\documentstyle[11pt,aaspp4]{article}
\def\mic{$\mu$m\ }
\def\nwat{nW m$^{-2}$ sr$^{-1}$}
\begin{document}
\small
\title{A TENTATIVE DETECTION OF THE COSMIC INFRARED BACKGROUND
AT 3.5 \mic FROM {\it COBE}/DIRBE OBSERVATIONS}
\author{E. Dwek\altaffilmark{1}, R. G. Arendt\altaffilmark{2}}

\altaffiltext{1}{Laboratory for Astronomy and Solar Physics, Code 685,
NASA/Goddard
Space Flight Center, Greenbelt, MD 20771.\ \ e-mail address:\
eli.dwek@gsfc.nasa.gov}
\altaffiltext{1}{Raytheon STX, Code 685, NASA/Goddard
Space Flight Center, Greenbelt, MD 20771}

\begin{abstract}
Foreground emission and scattered light from interplanetary dust (IPD)
particles and emission from
Galactic stellar sources are the
greatest obstacles for determining the cosmic infrared background (CIB)
from diffuse sky
measurements in the $\sim$ 1 to 5 \mic range. We use ground$-$based
observational limits on the K$-$band intensity of the CIB in conjunction
with skymaps obtained
by the Diffuse Infrared Background Experiment (DIRBE) on the 
{\it Cosmic Background Explorer} ({\it COBE})
satellite to reexamine the limits on the CIB at 1.25, 3.5, and 4.9 $\mu$m.
Adopting a CIB intensity of 7.4 \nwat at 2.2 $\mu$m, 
and using the 2.2 \mic DIRBE skymap
from which the emission from IPD cloud
has been subtracted, we create a spatial template of the Galactic
stellar contribution to the diffuse infrared sky. This template is then
used to subtract
the contribution of the diffuse Galactic stellar emission from the 
IPD--emission--subtracted DIRBE skymaps
at 1.25, 3.5, and 4.9 $\mu$m. The DIRBE 100 $\micron$ data are used to estimate
the small contribution of emission from interstellar dust
 at 3.5 and 4.9 $\micron$.
Our method significantly reduces the errors associated with the
subtraction of Galactic starlight, leaving only the IPD emission
component as the primary obstacle for the detection of the CIB at
these wavelengths.
The analysis leads to a tentative detection of
the CIB at 3.5 \mic with an intensity of
$\nu I_{\nu} = \{9.9 + 0.312[\nu_0 I_{CIB}(\nu_0) -\ 7.4]\} \pm 2.9$
\nwat, where
$\nu_0 I_{CIB}(\nu_0)$ is the CIB intensity at 2.2 \mic in units of \nwat.
The analysis also yields new upper
limits (95\% CL) on the CIB at 1.25 and 4.9 \mic of 68 and 36 \nwat, 
respectively.
The cosmological implications of these results are discussed in the paper.

\end{abstract}


%*************************************************************************
\section{INTRODUCTION}

Determination of the cosmic infrared background (CIB) from diffuse sky
measurements is
 greatly hampered by the presence of foreground emission and scattered
light from the
interplanetary dust (IPD) cloud, and emission 
from discrete and unresolved stellar components in our Galaxy, and from 
dust in the interstellar medium (ISM).
In a recent publication, Hauser et al. (1998) presented the results of the
search for the
 (CIB) in the 1.25 to 240 \mic wavelength region that was conducted with
the Diffuse Infrared Background Experiment (DIRBE)
 on the {\it Cosmic Background Explorer} ({\it COBE}) satellite. Careful
subtraction of foreground emission from the
IPD cloud
(Kelsall et al. 1998) and from stellar and interstellar Galactic emission
components (Arendt et al. 1998) revealed a residual
emission component in the DIRBE skymaps that, after detailed analysis of the
random and systematic uncertainties, was consistent with
 a positive signal at 100, 140, and 240~$\mu$m. Subsequent rigorous tests
showed that only the 140 and 240 \mic signals were
isotropic, a strict requirement for their extragalactic origin. Only upper
limits for the CIB intensity were given for
$\lambda$~=~1.25~$-$~60~$\mu$m, where the CIB
detection was hindered by residual emission from the IPD cloud. In the 1.25
to 4.9~\mic wavelength region,
 uncertainties in the subtraction of the Galactic stellar component
contributed to the uncertainties as well. The upper limits on the CIB
determined by Hauser et al. (1998) can be found in Table 2.

In this paper we use a new empirical model for the Galactic stellar emission in
order to reduce the uncertainties attributed by the DIRBE team to the
removal of this 
component. Our analysis relies on the modeling and error analysis of
the IPD and the ISM contributions to the diffuse
emission presented by Kelsall et al. (1998) and Arendt et al. (1998).
We use the DIRBE 1.25, 2.2, 3.5, and 4.9 \mic all sky maps from which the
emission from interplanetary
(zodiacal) dust
 has been subtracted (Kelsall et al. 1998) as a starting point in our
analysis. The intensity, $I_G(\lambda_j)$, of
 these maps should, in principle, contain only
the Galactic emission (starlight and ISM emission) and the CIB.
Assuming that the CIB is somehow determined from other independent
observations at wavelength $\lambda_0$, then
 subtraction of this background, $I_{CIB}(\lambda_0)$, from the corresponding
DIRBE skymap will create a spatial template of the Galactic
emission at this wavelength. 
At $\lambda_0\ =\ 2.2\ \micron$, emission from the ISM is negligible, and this 
template corresponds to the Galactic starlight.
Correlation of this stellar template map, $S_{\nu}(\lambda_0)$, with
the zodi$-$subtracted skymaps should yield
a straight line with a slope of
$S_{\nu}(\lambda_j$)/$S_{\nu}(\lambda_0$) 
corresponding to the
average Galactic stellar flux ratio,
and a non$-$zero intercept at zero $S_{\nu}(\lambda_0)$ intensity, which one can
identify as the CIB contribution to $I_G({\lambda_j})$ after a small correction
for the ISM emission is subtracted at 3.5 and 4.9 $\micron$.

The method is based on the assumption that the difference map $I_G(\lambda_0) -
I_{CIB}(\lambda_0)$ is a good representation of $S_{\nu}(\lambda_0)$, the intensity map 
of the Galactic stellar contribution at $\lambda_0$. 
Correlation of $S_{\nu}(\lambda_0)$ with maps at other wavelengths
will allow the identification and removal of the stellar emission at those wavelengths.
For $I_{CIB}(\lambda_0)$ we use the recently determined ground$-$based K$-$band
($\lambda=2.2~\micron$) galaxy counts, which give
 a strict lower limit of 7.4 \nwat\ to the CIB intensity at that wavelength
(e.g. Gardner 1996). Plots of the K-band contribution to the CIB
versus the magnitude (e.g. Pozzetti et al. 1998) show a decrease in the intensity at
faint magnitudes, suggesting that the contribution of faint
galaxies to the integrated light is small. We therefore adopt the
above intensity as the nominal value for the CIB at 2.2 $\micron$, and
explicitly state the dependence of our results on this value.

We apply the method to regions in the DIRBE skymaps that were identified in
the DIRBE analysis as high quality (HQ) regions where errors in
subtractions of the foreground emission are expected to be small
 (Hauser et al. 1998; see also \S2.1 below). The results show an extremely
good correlation between $S_{\nu}(2.2~\mu$m) and $S_{\nu}(\lambda_j)$
($\lambda_j$ = 1.25, 3.5, and 4.9 $\mu$m), characterized by a narrow
dispersion and a positive intercept.
Concentrating on HQB, the high quality region emphasized in the analysis by the
DIRBE team, we identify the
 uncertainties in the method as the difference between the intercept in
HQBN and HQBS, the northern and southern sections
 of HQB. These uncertainties are small compared with those associated with
the subtraction of the Faint Source Model (FSM) which the DIRBE team had
 used to characterize the contribution of unresolved
Galactic stellar sources (Arendt et al. 1998). At 1.25
and 4.9~$\micron$, uncertainties in the residuals are about equally
divided
 between the systematic errors in the subtraction of the zodiacal
light and the subtraction of the FSM 
contributions to the emission.
So a priori we do not expect our method to result in a significant
reduction in the uncertainties in the CIB determination at these wavelengths.
 However, at 3.5 $\micron$, uncertainties in the Hauser et al. (1998) 
results are dominated by
systematic errors in the FSM, which are significantly
larger than the errors in the proposed method. With this reduction of
uncertainties, 
our analysis yields a positive detection of the CIB at 3.5 
$\mu$m. The values derived for the CIB in different regions 
of the sky falls within the uncertainties, providing a simple 
test for their isotropic character.

%*************************************************************************
\section{ANALYSIS}

\subsection{An Empirical Model of Galactic Starlight}

As described in the Introduction, we adopt a value of $\nu I_{CIB}(2.2~
\mu$m) = 7.4 \nwat\ for the CIB intensity in
the K$-$band, and create a difference map $S_{\nu}(2.2~\mu$m) $\equiv$ \
$I_G(2.2~\mu$m) $-\ I_{CIB}(2.2~\mu$m), which
we identify as the spatial template of the Galactic stellar contribution to
the diffuse IR sky.
We chose for our analysis high quality (HQ) regions of the sky, identified
as such for their location at high Galactic latitudes ($b$)
and ecliptic latitudes ($\beta$), in which the contributions from the 
Galactic and zodiacal
emission components are relatively small compared to
other regions of the sky. HQA is a region encompassing 8780 deg$^2$ of
the sky at {$|b| > 30\arcdeg, |\beta| > 
25\arcdeg$}, including sections in both the northern
and southern hemispheres. The HQB region is a smaller area of 854
deg$^2$ contained within HQA at {$|b| > 60\arcdeg, |\beta| > 
45\arcdeg$}. Both HQ regions exclude several locations where the
ISM intensity is strong, $I_{ISM} > 0.2$ MJy/sr.

Figure 1 shows the correlation of $\nu I_G(\lambda_j)$ versus $\nu S_{\nu}(2.2~
\mu$m) within HQB for $\lambda_j$ = 1.25, 3.5, and 4.9 $\mu$m. 
A linear least--squares fit was performed to derive the intercepts and 
slopes of these correlations in addition to the correlation coefficients.
Results of these fits for the HQA and HQB regions and their north and south 
subregions (e.g. HQAN, HQAS) are presented in Table 1.
 The slopes of the correlations in Figure 1 and Table 1 represent the colors of
stellar emission. Most of this emission is unresolved (the brightest resolved 
sources lie outside the ranges displayed in Figure 1 and were excluded 
from the fit), but the colors are still a very good match to those of M and K
giants as shown in Figure 2 of Arendt et al. (1994). The intercepts of the 
correlations, $I_{int}(\lambda_j)$, give the remaining intensity of
any emission unassociated with 
Galactic starlight. At 3.5 and 4.9~$\micron$ this includes emission from 
the ISM. Following Arendt et al. (1998), we estimate and 
remove the mean level of this ISM 
emission by scaling the average intensity of the 100 $\micron$ ISM emission
seen by DIRBE in these regions
by factors of 0.00183 and 0.00292 at 3.5 and 4.9~$\micron$
respectively.
The resulting contribution of the ISM to the emission in each band is
also listed in Table 1.

\begin{figure}[h]
\epsscale{0.5}
\plotone{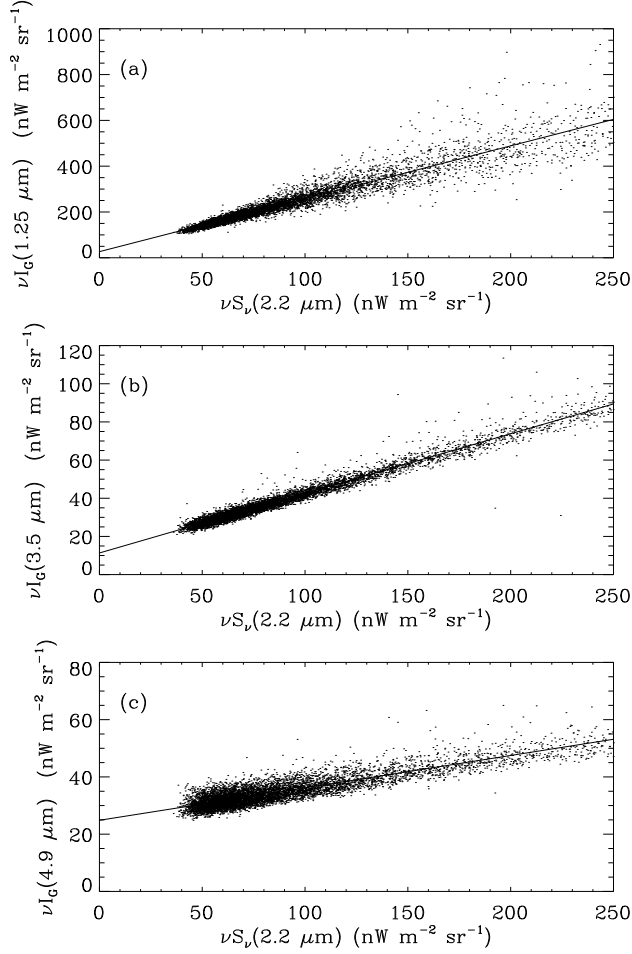}
\caption{Correlation plots of $\nu S_{\nu}(2.2~\micron)$ vs.
(a) $\nu I_{G}(1.25~\micron)$, (b) $\nu I_{G}(3.5~\micron)$, 
and (c) $\nu I_{G}(4.9~\micron)$ for the HQB region. The straight lines
are linear least--squares fits to the data with parameters as indicated
in Table 1.}
\end{figure}

%---------------------------
\subsection{Uncertainties}
%---------------------------
Three sources of errors contribute to the uncertainties in the derived
intensity of the CIB. 
The first consists of the uncertainties in the
residuals of the zodiacal emission, $\sigma_Z$, contained in the DIRBE
skymaps $I_G(\lambda_j)$, ($\lambda_j$ = 1.25, 3.5, and 4.9 $\mu$m),
 as well as the $S_{\nu}(2.2~\mu$m) difference map. 
The second uncertainty is that associated with the subtraction of the 
emission from the ISM. The third results from the
uncertainty in the value of the intercept, denoted
hereafter as $\sigma_{int}$.
The total uncertainty in the residual emission is given by:
\begin{eqnarray}
\sigma_{res}(\lambda_j)& = & \left[ \sigma_Z^2(\lambda_j)
			+ \sigma_{ISM}^2(\lambda_j) 
                        + [S_{\nu}(\lambda_j)/S_{\nu}(2.2~\mu m) 
                          \sigma_Z(2.2~\micron)]^2
                        + \sigma_{int}^2(\lambda_j) \right]^{1/2}  
                          \nonumber \nl
                      & \equiv & \left[ \sigma_Z^2(\lambda_j)
			+ \sigma_{ISM}^2(\lambda_j) 
                        + \sigma_{star}^2(\lambda_j)\right]^{1/2} 
\end{eqnarray}
\noindent
where $\sigma_{star}(\lambda_j)\equiv [S_{\nu}(\lambda_j)/S_{\nu}(2.2~\mu m) 
                          \sigma_Z(2.2~\micron)]^2
                        + \sigma_{int}^2(\lambda_j)$, is the total
uncertainty associated with our empirical stellar model. 
Table 2 lists the total uncertainty, $\sigma_{res}(\lambda_j)$, and
the various contributions to the uncertainty in the value of
$I_{res}(\lambda_j)$. The values of $\sigma_Z(\lambda_j)$ and 
$\sigma_{ISM}(\lambda_j)$ were taken from Arendt et al. (1998; Table 6), and
the value of $\sigma_{int}(\lambda_j)$ was
taken to be equal to the absolute value of the difference in the value of
$I_{int}(\lambda_j)$ between in HQBN and HQBS.
Also listed in the table are the systematic errors,
$\sigma_{FSM}(\lambda_j)$, derived by Arendt et al. (1998; Table 6) due to
uncertainties in the subtraction of the FSM. Comparison of 
$\sigma_{star}(\lambda_j)$ to
the error associated with the FSM illustrates the effectiveness of our
method in reducing
the uncertainties in removing the Galactic stellar emission component.

%----------------------------------
\subsection{Residual Intensities}
%----------------------------------

The residual intensity after subtraction of the IPD, stellar, and ISM 
foregrounds, is given by: 
$I_{res} = I_{int} - I_{ISM}$.
Following Hauser et al. (1998) we derive our results from the analysis of
the HQB region, which can be summarized as:
\begin{eqnarray}
\nu I_{res} ({\rm nW\ m}^{-2} {\rm\ sr}^{-1}) & = & 26.9 + 2.31 \ 
\Delta_{CIB}(2.2~\mu \rm m) \pm 20.6 \quad 
{\rm for\ } \lambda = 1.25\ \mu{\rm m} \nonumber \nl
       & = & 9.9 + 0.312\ \Delta_{CIB}(2.2~\mu \rm m) \pm 2.9  \quad
{\rm for\ } \lambda = 3.5\ \mu{\rm m} \nl
       & = & 23.3 + 0.113\ \Delta_{CIB}(2.2~\mu \rm m) \pm 6.4 \quad 
{\rm for\ } \lambda = 4.9\ \mu{\rm m} \nonumber
\end{eqnarray}
\noindent
where the quoted 
errors represent 1 $\sigma$ uncertainties, and 
$\Delta_{CIB}(2.2~\mu \rm m) \equiv [I_{CIB}(2.2\ \mu{\rm m}) - 7.4]$
represents the difference in the actual value of the CIB at
2.2~$\micron$ 
(in \nwat) and the nominal value adopted in our model. Values of $\nu
I_{res}(\nu)$ for $\Delta_{CIB}(2.2~\mu \rm m)=0$ are listed in
Table~2. Only the 3.5 and 4.9 $\micron$ residuals are positive ($> 3 \sigma$) 
and therefore potential detections of the CIB.  

%--------------------------
\subsection{Isotropy Tests}
%--------------------------
We tested the residual
intensities in the 3.5 and 4.9 $\micron$ bands for isotropy, an
expected property of the CIB.
Table 1 shows that the residual 3.5~$\micron$ emission is isotropic,
as evident from
the agreement between the mean intensities in the
various HQ regions. However, a more detailed analysis shows the
presence of a low$-$level large scale gradient. This gradient is
caused by residual emission from IPD particles. This gradient would
have been undetected in the analysis of Hauser et al. (1998) using the
FSM to subtract the Galactic stellar emission. The reduction in this
gradient from the level found by Hauser et al. (1998) to its present
level is the result of our more accurate subtraction of the
stellar emission component. 
Without the existence of this weak gradient, the residual would
classify as a definite detection of the CIB. Instead, we
conservatively regard this result as a tentative detection of the CIB
at 3.5~$\micron$.

The same gradient seen in the 3.5 $\micron$ image is significantly 
stronger in the
4.9 $\micron$ residual map, and is responsible for the larger
dispersion in the mean intensities in the various HQ regions. We
therefore consider the 4.9 $\micron$ result as only an upper limit.

\begin{figure}[h]
\epsscale{0.5}
\plotone{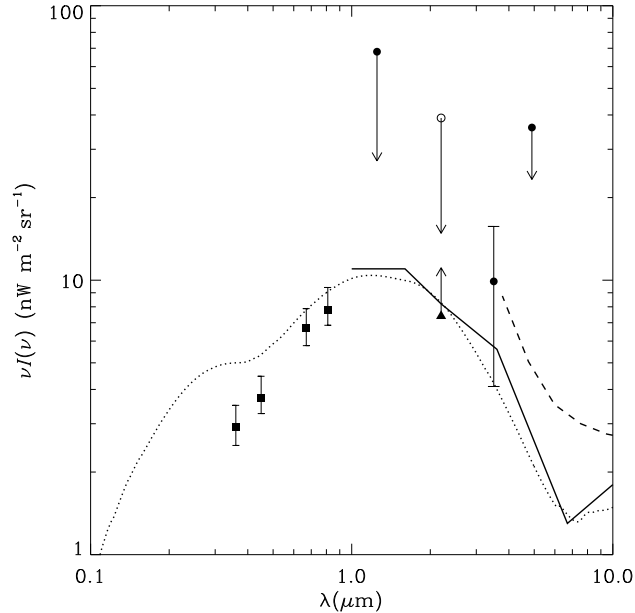}
\caption{UV determination of the EBL and
our {\it COBE}/DIRBE limits and tentative 3.5 $\micron$
detection (solid circles)
 are compared to select model predictions: Franceschini et al. (1997;
solid line), Malkan \& Stecker (1998; dashed line). The dotted line
represents model
$ED$ from Dwek et al. (1998) scaled up by a factor of
1.2. The solid squares
represent HDF detections by Pozzetti et al. (1998), the solid
triangle represents the 2.2 $\micron$ lower limit from galaxy counts (Gardner
1996), and the open circle is the DIRBE upper limit at 2.2~$\micron$
(Hauser et al. 1998).}
\end{figure}

%*************************************************************************
\section{DISCUSSION}
%*************************************************************************

Figure 2 shows the current limits on and tentative detection of the
extragalactic background light (EBL) in the 0.1 to 10 \mic wavelength
region. The filled squares represent the detections at 0.36,
0.45, 0.67, and 0.81 $\micron$ from background fluctuations analysis of the
Hubble Deep Field (Pozzetti et al. 1998). The total EBL intensity
in the 0.36 to 3.5 $\micron$ wavelength region is  16 \nwat, about
equal to the total integrated CIB intensity in the 140 to 1000
$\micron$ region (Hauser et al. 1998; Fixsen et al. 1998; Dwek et al. 1998).

Figure 2 compares several model predictions to the observational
constraints.
The solid line represents the model of Franceschini et
al. (1997), and the dashed line represents that calculated by 
Malkan \& Stecker (1998). 
Both models are consistent with the tentative
3.5 $\micron$ detection. However, they are incomplete with regard to
their
 EBL prediction at shorter wavelengths. The dotted line represents two models
calculated by Dwek et
al. (1998) in which the UV and optically determined cosmic star
formation rate presented by Madau et al. (1998) was augmented by
 hidden starbursts at either $z \approx 1.5$ (model $ED$) or by
hidden starbursts at high redshifts (model $RR$). The models give
similar results in the 0.1 to 10 $\micron$ wavelength regions, and were
scaled up by a factor of $1.2$ in order to provide a better fit
to the tentative 3.5 $\micron$ detection. The models produce however
an excess flux at 0.36 and 0.45 $\micron$ requiring some
modifications in the model assumptions.

Most models to date have concentrated on predicting the EBL at mid$-$
and far$-$IR wavelengths, and
appear to be incomplete in the UV and near$-$IR regions of the
spectrum. Our tentative detection of the CIB at 3.5 $\micron$ adds
an important new constraint on CIB models in this wavelength region.

\acknowledgments
We acknowledge helpful conversations with Jonathan Gardner and Harvey
Moseley, and thank Janet Weiland, Tom Kelsall, and Dave Leisawitz for
their 
useful comments on the manuscript.
This research was supported by the NASA Astrophysical Theory Program 
NRA97-12-OSS-098.

%*************************************************************************

\clearpage

%Table 1
\begin{deluxetable}{ccccccc}
\tablecaption{Results of Correlations Identifying Emission Components}
\footnotesize
\tablehead{
\colhead{Region}&
\colhead{Wavelength}&
\colhead{Number of}&
\colhead{Correlation}&
\colhead{Slope =} &
\colhead{Intercept = $\nu I_{int}(\lambda_j)$}&
\colhead{$\nu I_{ISM}(\lambda_j)$}\\
\colhead{}&
\colhead{($\micron$)}&
\colhead{pixels}&
\colhead{Coefficient}&
\colhead{${\nu S_{\nu}(\lambda_j)\over \nu S_{\nu}(2.2~\micron)}$} &
\colhead{(nW m$^{-2}$ sr$^{-1}$)}&
\colhead{(nW m$^{-2}$ sr$^{-1}$)}
}
\startdata
{\bf HQB} & {\bf 1.25} &  {\bf 7460} & {\bf 0.95} & {\bf 2.310 $\pm$ 0.009} & {\bf 26.9 $\pm$ 0.8} & \nodata \nl
HQBN & 1.25 &  3643 & 0.95 & 2.315 $\pm$ 0.012 & 28.0 $\pm$ 1.1 & \nodata \nl
HQBS & 1.25 &  3817 & 0.95 & 2.311 $\pm$ 0.013 & 25.3 $\pm$ 1.3 & \nodata \nl
HQA  & 1.25 & 73917 & 0.96 & 2.288 $\pm$ 0.002 & 30.4 $\pm$ 0.3 & \nodata \nl
HQAN & 1.25 & 37353 & 0.96 & 2.302 $\pm$ 0.003 & 30.7 $\pm$ 0.4 & \nodata \nl
HQAS & 1.25 & 36564 & 0.96 & 2.280 $\pm$ 0.004 & 29.6 $\pm$ 0.4 & \nodata \nl
\hline

{\bf HQB} & {\bf 3.5} &  {\bf 7461} & {\bf 0.98} & {\bf 0.3123 $\pm$ 0.0007} & {\bf 11.34 $\pm$ 0.07} & {\bf 1.4} \nl
HQBN & 3.5 &  3643 & 0.98 & 0.3112 $\pm$ 0.0010 & 10.93 $\pm$ 0.09 & 1.2 \nl
HQBS & 3.5 &  3818 & 0.98 & 0.3111 $\pm$ 0.0009 & 11.93 $\pm$ 0.09 & 1.5 \nl
HQA  & 3.5 & 73905 & 0.98 & 0.3141 $\pm$ 0.0002 & 11.95 $\pm$ 0.03 & 2.2 \nl
HQAN & 3.5 & 37349 & 0.98 & 0.3125 $\pm$ 0.0003 & 11.69 $\pm$ 0.04 & 2.1 \nl
HQAS & 3.5 & 36556 & 0.98 & 0.3142 $\pm$ 0.0003 & 12.36 $\pm$ 0.04 & 2.3 \nl
\hline

{\bf HQB} & {\bf 4.9} &  {\bf 7461} & {\bf 0.88} & {\bf 0.1133 $\pm$ 0.0007} & {\bf 24.85 $\pm$ 0.07} & {\bf 1.6} \nl
HQBN & 4.9 &  3643 & 0.91 & 0.1116 $\pm$ 0.0008 & 23.89 $\pm$ 0.08 & 1.4 \nl
HQBS & 4.9 &  3818 & 0.88 & 0.1101 $\pm$ 0.0010 & 26.19 $\pm$ 0.10 & 1.7 \nl
HQA  & 4.9 & 73901 & 0.66 & 0.0987 $\pm$ 0.0004 & 28.31 $\pm$ 0.05 & 2.5 \nl
HQAN & 4.9 & 37348 & 0.72 & 0.1015 $\pm$ 0.0005 & 26.96 $\pm$ 0.06 & 2.4 \nl
HQAS & 4.9 & 36553 & 0.61 & 0.0922 $\pm$ 0.0006 & 30.10 $\pm$ 0.07 & 2.6 \nl
\enddata
\end{deluxetable}

%Table 2
\begin{deluxetable}{ccccc}
\tablecaption{Residual Emission and Uncertainties\tablenotemark{a}}
\tablehead{
\colhead{$\lambda(\mu$m)} &
\colhead{1.25}  &
\colhead{2.2}  &
\colhead{3.5} &
\colhead{4.9}
}
\startdata
$\sigma_Z$      &    15.     &   6.    & 2.          &  6.    \nl
$\sigma_{ISM}$  & \nodata    & \nodata & 0.4         &  0.6   \nl
$\sigma_{int}$  &     2.7    & \nodata & 0.70        &  2.0   \nl
$\sigma_{res}$  &    20.6    & \nodata & 2.86        &  6.39  \nl
\hline
$\sigma_{star}$ &    14.1    & \nodata & 2.0         &  2.1   \nl
$\sigma_{FSM}$  &    15.     &   10.   & 6.          &  5.    \nl
\hline
$\nu I_{res}(HQB)$    & $26.6~\pm~20.6$    & \nodata & 9.9$~\pm$ 2.9 &
23.3 $\pm$ 6.4 \nl
Isotropy         &    no  &  assumed  & weak  & no \nl
\hline
$\nu I_{CIB}$\tablenotemark{b}  &  $<68 $  &  7.4 (assumed) & $9.9~\pm~2.9$  &  $< 36$

\enddata
\tablenotetext{a}{All values are given in units of \nwat.}
\tablenotetext{b}{For comparison, $\nu I_{CIB}$ intensities from
Hauser et al. (1998) are: $<75$, $<39\ (14.9 \pm 12)$, $<23\ (11.4~\pm~6)$, and $<44$ \nwat at 1.25, 2.2,
3.5, and 4.9 $\micron$, respectively.} 
\end{deluxetable}

%*************************************************************************

\begin{thebibliography}{}
\bibitem[ ]{MAr78} Aaronson, M. 1978, ApJ, 221, L108
\bibitem[ ]{Are94} Arendt, R. G. et al. 1994, \apj, 425, L85
\bibitem[ ]{Are98} Arendt, R. G. et al. 1998, \apj, in press (astro-ph/9805323)
\bibitem[ ]{Dwe98} Dwek, E. et al. 1998, \apj, in press (astro-ph/9806129)
\bibitem[ ]{Fix98} Fixsen, D. J., Dwek, E., Mather, J. C., Bennett, C. L., \&
                   Shafer, R. A. 1998, \apj, in press (astro-ph/9803021)
\bibitem[ ]{Fra97} Franceschini, A. et al. 1997, in The Far InfraRed
                   and Submillimetre Universe, ed. A. Wilson 
                   (Noordwijk: ESA Publication Division), 159
\bibitem[ ]{Gar95} Gardner, J. P. 1996, in Unveiling the Cosmic Infrared
                   Background, ed. E. Dwek (New York: AIP Press), p. 127
\bibitem[ ]{Hau98} Hauser, M. G. et al. 1998, \apj, in press (astro-ph/9806167)
\bibitem[ ]{Kel98} Kelsall, T. J., et al. 1998, \apj, in press 
                   (astro-ph/9806250)
\bibitem[ ]{Mad97} Madau, P., Pozzetti, L., \& Dickinson, M. 1998,
                   \apj, 498, 106
\bibitem[ ]{Mal97} Malkan, M. A., \& Stecker, F. W. 1998, \apj, 496, 13
\bibitem[ ]{Poz97} Pozzetti, L., Madau, P., Ferguson, H. C., Zamorani, G., \&
                   Bruzual, G. A. 1998, \mnras, in press (astro-ph/9803144)
\end{thebibliography}
\end{document}